\documentclass{article}
\usepackage[left=3cm,top=2.5cm,right=3cm,bottom=2.5cm]{geometry} 

\usepackage{amsmath}
\usepackage{mathrsfs}
\usepackage{slashbox}
\usepackage{multirow}
\usepackage{epsfig}
\usepackage{enumerate}

\title{Recurrence in resonant transmission of one-dimensional array of delta potentials}
\author{G. Cordourier-Maruri\mbox{$^{1,2}$}, \ \ V. Gupta\mbox{$^{1}$} \ \ and \ \ R. de Coss\mbox{$^{1}$} \\ 
\mbox{$^{1}$} Departamento de F{\'i}sica Aplicada Cinvestav-M\'{e}rida,\\
A.P. 73 Cordemex 97310 M\'{e}rida, Yucat\'an, M\'{e}xico. \\
\mbox{$^{2}$}Department of Physics and Astronomy, University College London, \\
Gower Street, London WC1E 6BT, United Kingdom.
}

\begin{document}
\large
\maketitle

\begin{abstract} 
\large
The resonant transmission of a moving particle which interacts with an one-dimensional array of $N$ $\delta$-function potentials is investigated. A suitable
transfer matrix formulation is used to obtain the particle transmission. We give the parameters for perfect tunnelling and the transcendental equation for
the quasi-bound state energies for $N = 2$, $3$ and $4$. Conditions for perfect tunnelling and resonant transmission are discussed for arrays with arbitrary 
$N$. A model to explain how the tunnelling energy filter works in these systems is proposed here.
\end{abstract}

\section{Introduction} \label{intro}

The property of resonant transmission also known as resonant tunnelling, is an interesting topic from the point of view of both the physical understanding and
the practical applications of potential barrier arrays. It has been extensively studied in different fields of physics involving wave 
propagation\cite{gri,pw}. The resonant transmission is present in mechanical waves analyzed in acoustic\cite{cra,san}, in electromagnetic waves studied in
optics\cite{pochi,ben,va}, and in quantum mechanics\cite{gri2,wo,spr,cv,bra}. From Esaki seminal papers\cite{esa1,esa2,esa3}, electronic perfect tunnelling
in a double barrier system has been explained in terms of the quasi-bound states (QBS) between the barriers\cite{esa4}. This understanding allowed the
fabrication of resonant tunnelling diodes (RTD), which present negative differential resistance and good performance in fast processes\cite{esa3,ngu}.
Electronic resonance is also present with three\cite{esa2} and more potential barriers. The specific conditions to obtain resonance depends on recurrences
in the transmission functions, and on the shape of the barrier array, which can be divided into subsets of potentials or cells\cite{gri,spr}.  

In this work we consider $\delta$-function shaped potential barriers. Such a potential array has been useful to model several solid state systems like
magnetic impurities and short range interactions\cite{ro1,ro2,ro3,ro4}. It has been used to study the conduction properties of crystals through the 
Kronig-Penney model\cite{kit} and Anderson localization in a disordered impurity array\cite{and,hue,koh,si}. In a solid state quantum information scenario,
$\delta$-function potential barriers are used to depict the instantaneous interaction between a flying spin and a fixed magnetic impurity 
\cite{bos,cic1,cic2,cor2,cicn2,cicn}, to implement teleportation\cite{cit} and quantum memory\cite{qm}. 

Here we study the conditions under which an array of $N$ $\delta$-function potential barriers, gives resonant transmission with an incident particle. The
situation depicted here can be implemented, in a solid state scenario with a ballistic electron moving on a carbon nanotube\cite{white,balents}, a
heterostructure or a quantum Hall edge states\cite{hermelin,mcnell}, where short range potentials, impurities or quantum dots are located. 

This paper is organized as follows. In order to clarify the results and discussion, in the section 2 we give the transfer matrix method and its
representation which was given recently\cite{cor}. In section 3 we present the conditions for resonant transmission and perfect tunnelling for arrays
with $N=2$, $3$ and $4$. Here we also calculate the transcendental equations to find the QBS energies of these arrays. In section 4, we generalize the
results for arrays with arbitrary $N$ in terms of sub-arrays or cells. The intrinsic QBS energy concept helps us to explain the working of a tunnelling energy filtering in such arrays. Section 5 gives the conclusions.

\section{Transfer matrix}

\begin{figure}[h]
\begin{center}
\resizebox{14cm}{!}{\includegraphics{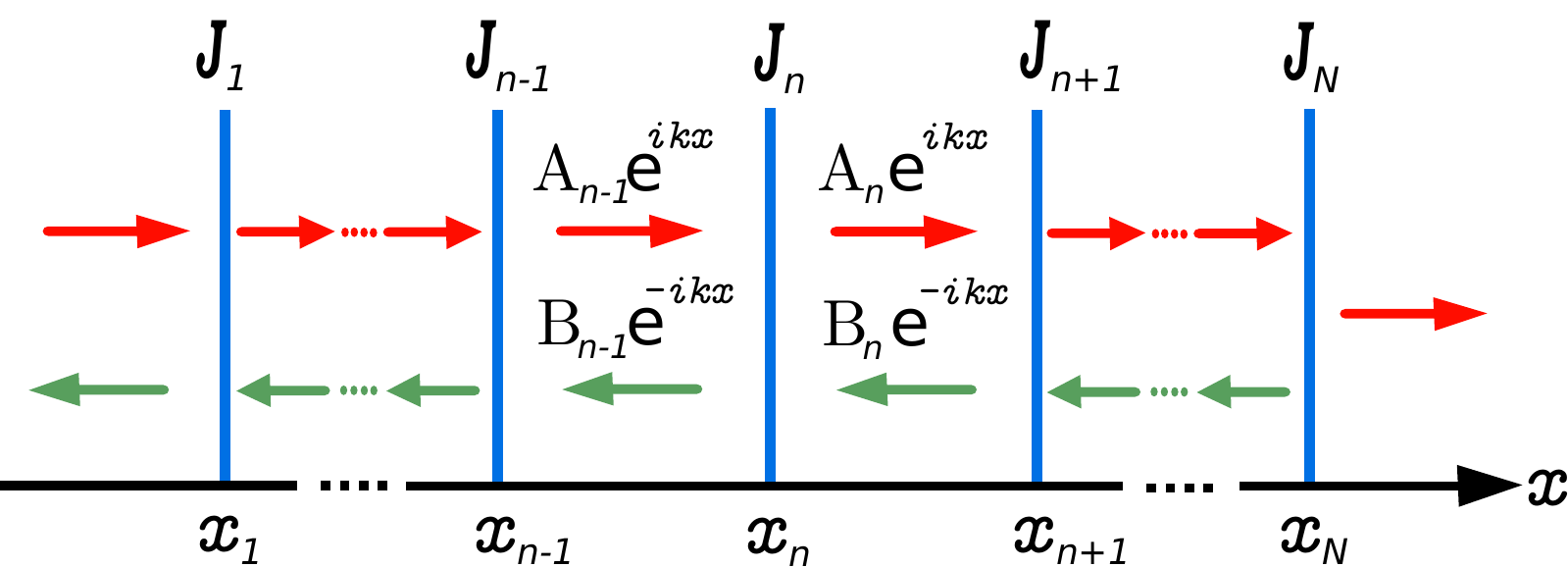}}
\caption{(Color online) Scheme of one dimensional quantum wire with $N$ $\delta$-function potential barriers of strength $J_n$ located at $x=x_n$, 
          $n=1,2,..,N$. A moving particle (with wave number $k$) incident from the left, is scattered off the potential array. The transmitted (or reflected)
          particle is indicated by right (left) pointed arrows.}
\end{center}
\end{figure}

We consider an $N$ $\delta$-function potential barriers on a one-dimensional quantum wire along the $x$-axis. A moving particle with energy $\epsilon$, is
incident from the left end of the wire, as is shown in Figure 1. The Hamiltonian of the system is

\begin{equation}
\label{h1}
\hat{H} =\frac{p^2}{2m} + \sum_{n=1}^NJ_n\delta(x-x_n),
\end{equation} 

\noindent
where $p$ and $m$ are the momentum and mass of the particle, respectively. $J_n$ is the strength and $x_n$ is position of the $n$-th potential barrier. The
particle wavefunction $\psi_n(x)$ in the region $x_{n}<x<x_{n+1}$, is taken to be

\begin{equation}
\psi_n(x) = A_n e^{ikx} + B_n e^{-ikx},
\end{equation}

\noindent
where the wave number is $k=\sqrt{2m\epsilon}/\hbar$, and $A_n$ and $B_n$ are the probability amplitudes for incoming ($k$) and outgoing ($-k$) parts of the
wavefunction, respectively. To relate the coefficients ($A_{n-1}$, $B_{n-1}$,$A_n$ and $B_n$) of the wavefuction on both sides of the $n$-th potential
(see Figure 1), we use the transfer matrix ${\bf M}_n$

\begin{equation}
\left(\begin{array}{c} A_n \\ B_n \end{array}\right)=
\frac{1}{2}\left( \begin{array}{cc}
2-\imath \lambda_n  &  -\imath \lambda_n E_n^*  \\
\imath \lambda_n E_n & 2+\imath \lambda_n \end{array} \right)
\left(\begin{array}{c} A_{n-1} \\ B_{n-1} \end{array}\right)={\bf M}_n\left(\begin{array}{c} A_{n-1} \\ B_{n-1} \end{array}\right),
\ \ \ \ n=1,2...,N.
\end{equation}    

\noindent
in terms of the dimensionless strength parameter $\lambda_n = 2mJ_n/k\hbar^2$ and $E_n = \exp(2ikx_n)$. Another representation for the transfer matrix 
${\bf M}_n$ is

\begin{equation}
\label{d2}
{\bf M}_n = I-\frac{i \lambda_n}{2}{\bf L}_n    
\end{equation} 

\noindent
where

\begin{equation}
\label{d3}
      {\bf L}_n(E_n) =  \left( 
\begin{array}{cc}
                  1              &  E_n^*  \\
                  -E_n             & -1              
\end{array} \right).  
\end{equation} 

\noindent
This representation is useful to find the transfer matrix for $N$ potentials, defined as 
${\bf \mathcal{M}}(N)\equiv{\bf M}_N{\bf M}_{N-1}...{\bf M}_2{\bf M}_1$. ${\bf \mathcal{M}}(N)$ will relate the incident with the transmitted wavefuction, 
that is 

\begin{equation}
          \left(\begin{array}{c} A_N \\ B_N \end{array}\right) \equiv {\bf \mathcal{M}}(N)
          \left(\begin{array}{c} A_{0} \\ B_{0} \end{array}\right),
\end{equation} 

\noindent
where ${\bf \mathcal{M}}(N)$ can be expressed as\cite{cor}

\begin{align}\label{d1}
{\bf \mathcal{M}}(N)= I-\frac{i}{2}\sum_{n_1=1}^N\lambda_{n_1}{\bf L}_{n_1} + 
         \left(-\frac{i}{2}\right)^2\sum_{\substack{n_1,n_2=1\\n_1>n_2}}^N (\lambda_{n_1}{\bf L}_{n_1})(\lambda_{n_2}{\bf L}_{n_2}) +...
\nonumber       \\
...+\left(-\frac{i}{2}\right)^m\sum_{\substack{n_1,n_2,..,n_m=1\\n_1>n_2>...>n_m}}^N 
        (\lambda_{n_1}{\bf L}_{n_1})(\lambda_{n_2}{\bf L}_{n_2})...(\lambda_{n_m}{\bf L}_{n_m}) + 
\nonumber        \\
...+\left(-\frac{i}{2}\right)^N(\lambda_{n_1}{\bf L}_{n_1})(\lambda_{n_2}{\bf L}_{n_2})...(\lambda_{n_N}{\bf L}_{n_N}).
\end{align}

\noindent
This representation of ${\bf \mathcal{M}}(N)$ is very useful because of the interesting property of the $L$-matrices, namely

\begin{equation}
\label{con}
{\bf L}_n{\bf L}_m+{\bf L}_m{\bf L}_n = (2-E_nE_m^*-E_n^*E_m)I = 4\sin^2(\phi_{mn})I,
\end{equation} 

\noindent
where $\phi_{nm}=k(x_m - x_n)$. An obvious and useful consequence is that ${\bf L}_n^2=0$. Eq. (\ref{con}) is very useful in simplifying the multiple products
of ${\bf L}$'s in the expression for the general transfer matrix in Eq. (\ref{d1}) above.

In this way, for no particle incident from the right of the $N$-th potential, that is $B_N=0$, the probability of transmission is

\begin{equation}
       T = \frac{1}{|({\bf \mathcal{M}}(N))_{22}|^2},
\end{equation} 

\noindent
since det ${\bf \mathcal{M}}(N)=1$.

We now present results for resonant transmission and perfect tunnelling ($T=1$) for some specific arrays using the representation of ${\bf \mathcal{M}}(N)$ 
in Eq. (\ref{d1}).
     
\section{Results for $N = 2$, $3$ and $4$}

\medskip

In our previous work\cite{cor}, some specific regular arrays were considered. Here we note that $E_n = \exp(2ikx_n)$ is the square of the wave function at
$x=x_n$ for particle travelling to the right. If the distance between two potentials, at $x_n$ and $x_m$ is such that $k(x_n - x_m)=\pi\alpha_{nm}$ where 
$\alpha_{nm}$ is an integer then $E_n=E_m$, consequently ${\bf L}_n={\bf L}_m$.

If all the interpotential distances $k(x_n - x_m)$, ($n$, $m = 1$, $2$,...) are integer multiples of $\pi$ then $E_1=E_2=...=E_N$, implying 
${\bf L}_n={\bf L}$ for $n=1$, $2$, ..., $N$. Since ${\bf L}^2=0$, the transfer matrix in Eq. (\ref{d1}) reduces to \footnote{This will also hold for all wave
number $k'=\beta'k$ ($\beta$ an integer) since $k'(x_n - x_m) = \pi\beta\alpha$.}

\begin{equation}\label{mm}
       {\bf \mathcal{M}}(N)= I-\frac{i}{2}{\bf L}\sum_{n=1}^N\lambda_n.
\end{equation}  

\noindent
This means that the $N$ potential array effectively acts like a single potential of strength $\Lambda = \sum_{n=1}^{N}\lambda_n$. This behaviour is due to
resonance between the particle wavefuction and the array geometry. Note that the distance between any two potential locations need not be equal. The only
requirement is that all $k(x_n -x_m)$ be integer multiple of $\pi$. Clearly, a single potential can give $T=1$ only if it has zero strength. In the above
case this means $\sum_{n=1}^N\lambda_n=0$. 

\medskip

\subsection{Two-$\delta$-function potential array}
 
For $N = 2$ system Eq. (\ref{d1}) gives

\begin{equation}\label{dos}
       {\bf \mathcal{M}}(2)= I-\frac{i\lambda_2}{2}{\bf L}_2-\frac{i\lambda_1}{2}{\bf L}_1
       -\frac{\lambda_2\lambda_1}{4}{\bf L}_2{\bf L}_1.
\end{equation} 

\noindent
If $E_2=E_1$, then ${\bf L}_2={\bf L}_1$ and this reduces to an effective $N=1$ case. Thus for a genuine $N=2$ array we need ${\bf L}_2{\bf L}_1$, to be 
non-zero. From Eq. (\ref{dos}) one obtains

\begin{equation}
({\bf \mathcal{M}}(2))_{22} = \frac{1}{4}z_2z_1 + \frac{1}{4}\lambda_2\lambda_1E_2E_1^*,
\end{equation}

\noindent
where $z_n\equiv(2 + i\lambda_n)$, $n=$ $1$, $2$, ... . Thus, the transmission probability depends on three complex numbers namely $E_2E_1^*$, $z_1$ and 
$z_2$. The last two are contained in ${\bf M}_1$ and ${\bf M}_2$, respectively. The condition for perfect tunnelling ($T=1$), can be expressed as

\begin{equation}
E_2E_1^*=\pm \frac{z_1z_2}{|z_1z_2|}.
\end{equation}

\noindent
From this result, it is easy to see that if the two $\delta$-function potential have the same strength ($\lambda_1=\lambda_2=\lambda$), then for 
$E_2E_1=-z/z^*$ one obtains $T=1$ for any $\lambda$ (for graphical representation and other details see \cite{cor}). This can also be expressed as

\begin{equation}\label{qbs2}
\tan(2k(x_1-x_2))=\frac{4\lambda^2}{4+\lambda^2},
\end{equation} 

\noindent
where the wave number is $k=\sqrt{2m\epsilon}/\hbar$, and the parameter $\lambda= 2mJ_n/k\hbar^2=\sqrt{2mJ^2/\epsilon}/\hbar$. As the perfect tunnelling
is present only when the incident particle energy is equal to a QBS energy\cite{esa4}, Eq. (\ref{qbs2}) is the transcendental equation for the QBS energies
of an equal strength two-$\delta$-function potential array.

\medskip

\subsection{Three-$\delta$-function potential array}

For a $N = 3$ system Eq. (\ref{d1}) gives
                        
\begin{equation}\label{m3}
  {\bf \mathcal{M}}(3)= I-\frac{\imath}{2}\sum_{n_1=1}^3\lambda_{n_1}{\bf L}_{n_1}
                       -\frac{1}{4}\sum_{n_1,n_2=1}^3\lambda_{n_1}{\bf L}_{n_1}\lambda_{n_2}{\bf L}_{n_2}
                       +\frac{i}{8}\lambda_3\lambda_2\lambda_1{\bf L}_3{\bf L}_2{\bf L}_1.
\end{equation}

Using the properties of the ${\bf L}$-matrices in Eq. (\ref{con}), one can easily read off the following special cases.

\begin{enumerate}[i)]
  \item  $E_1=E_2=E_3$, that is ${\bf L}_1={\bf L}_2={\bf L}_3$ one has effectively a single $\delta$-function potential barrier of strength 
         ($\lambda_1+\lambda_2+\lambda_3$).

  \item For $E_1=E_2$ that is ${\bf L}_1={\bf L}_2$ or $E_2=E_3$, that is ${\bf L}_2={\bf L}_3$, one has an effective $N=2$ $\delta$-function potential array
        with strength ($\lambda_1 + \lambda_2$) and $\lambda_3$ or $\lambda_1$ and ($\lambda_2 + \lambda_3$). This is so because for these cases 
        ${\bf L}_3{\bf L}_2{\bf L}_1=0$ in Eq. (\ref{m3}).

  \item For a genuine $N=3$ array one needs the ${\bf L}_3{\bf L}_2{\bf L}_1$ term to be non-zero. In general, after a little algebra, Eq. (\ref{m3})
        gives 

        \begin{equation}\label{mm3}
        ({\bf \mathcal{M}}(3))_{22} = \frac{1}{8}[z_1z_2z_3 + \lambda_1\lambda_2z_3E_2E_1^* + \lambda_1\lambda_3z_2^*E_3E_1^* + \lambda_2
        \lambda_3z_1E_3E_2^*].
        \end{equation}

        \noindent
        Here $z_n = (2 + \imath\lambda_n)=2({\bf M}_n)_{22}$ for $n=1$, $2$ and $3$, while $E_n = \exp(2i\phi_n)$ with $\phi_n = kx_n$.
\end{enumerate}

The first term in Eq. (\ref{mm3}) represents the effect of single potential transfer matrices while the rest contain two independent relative phases from 
the wavefunction at the three sites of the delta function potential. For perfect tunnelling ($T=1$) these effects have to compensate each other. In other
words, for $T=1$ there will be specific relations between the strength parameters $\lambda_i$ (equivalently the phase of $z_i$) and the phase factors $E_i$
coming from the wavefunction.

For further analysis we set $E_1 = 1$. This is just a choice of the origin ($x_1=0$). Even so, Eq. (\ref{mm3}) depends on five complex variables. Guided by
the simplest non-trivial case of $N=2$, we take the phases of $z_i$ which contain the strength parameters, to be equal. That is, we take 
$\lambda_1=\lambda_2=\lambda_3=\lambda$. This reduces Eq. (\ref{mm3}) to 

\begin{equation}\label{m32}
({\bf \mathcal{M}}(3))_{22} = \frac{1}{8}(z^3 + \lambda^2zE_2 + \lambda^2z^*E_3 + \lambda^2zE_3E_2^*).
\end{equation}

\begin{figure}[h]
\begin{center}
\resizebox{16cm}{!}{\includegraphics{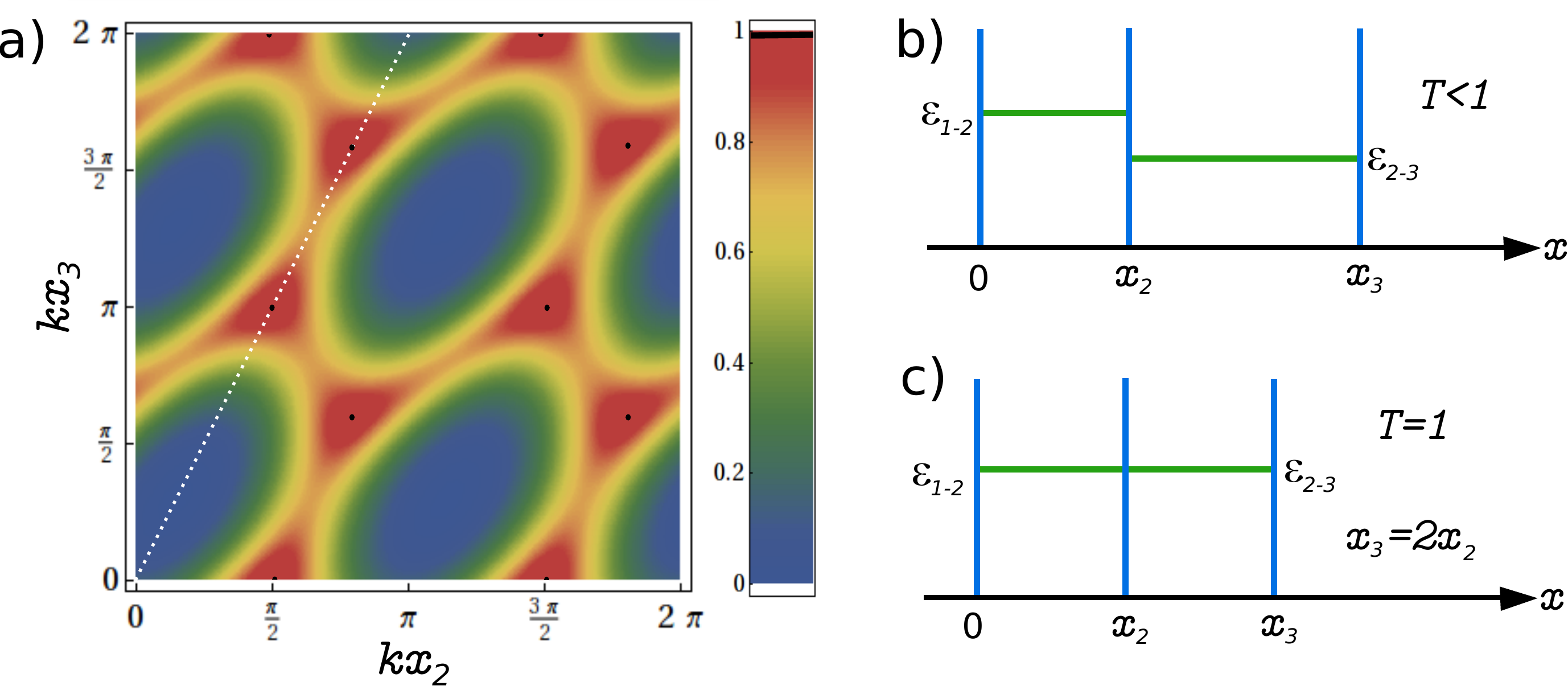}}
\caption{(Color online) Particle transmission for a genuine $N=3$ $\delta$-function potential array. a) Particle transmission $T$ as a function of $kx_2$ and
          $kx_3$, with $\lambda_1=\lambda_2=\lambda_3=1$, the white dotted line indicate the relation $kx_3=2kx_2$ and the black dots the perfect tunnelling.
          b) Scheme of an asymmetrical three $\delta$-function potential array, where the intrinsic QBS energies are different 
          ($\epsilon_{1-2} \neq \epsilon_{2-3}$) and $T<1$. c) Scheme of a symmetrical three $\delta$-function potential array, where the intrinsic QBS
          energies are equal ($\epsilon_{1-2} = \epsilon_{2-3}$) and we have perfect transmission.}
\end{center}
\end{figure}

The particle transmission ($T = 1/|({\bf \mathcal{M}}(3))_{22}|^2$) is plotted in Figure 2 a), as a function of $kx_2$ and $kx_3$, with 
$\lambda_1=\lambda_2=\lambda_3=1$. We can see the points of perfect tunnelling, which follows a $\pi$ periodicity in both $kx_2$ and $kx_3$. In addition,
notice that the perfect tunnelling points fulfil the relation $kx_3=2kx_2$. This behaviour can be explained if we consider the QBS energies in the 
three-potential array. Suppose that between every two contiguous $\delta$-function barriers, there is intrinsic QBS energy which depends on the
geometry of the two barriers. Then, consider an asymmetrical array, like the one depicted in Figure 2 b). It has different values for the intrinsic QBS
energies ($\epsilon_{1-2} \neq \epsilon_{2-3}$) and no perfect transmission is expected. By contrast, the symmetrical array described in Figure 2 c) presents
the same geometry for every two contiguous barriers and consequently the intrinsic QBS energies are equal ($\epsilon_{1-2}=\epsilon_{1-2}=\epsilon_{qbs}$). In
this case $\epsilon_{qbs}$ will be the QBS energy of the three-$\delta$-function barrier array. The necessary symmetry dictates that $x_3=2x_2$ or, given the
periodicity of the wavefunction, $x_3=2x_2+2\pi/k$. 

The last argument is incorporated in Eq. (\ref{m32}) by putting that $E_3=E_2^2$, so that

\begin{equation}\label{m33}
({\bf \mathcal{M}}(3))_{22} = \frac{1}{8}(z^3 + 2\lambda^2zE_2 + \lambda^2z^*E_2^2).
\end{equation}

\noindent
Imposing the perfect tunnelling condition, that is $T = 1/|({\bf \mathcal{M}}(3))_{22}|^2 = 1$, gives

\begin{equation}\label{qbs3}
\cos(2kx_2) = \frac{2 + \lambda^2 + 4\lambda\sin(2kx_2)}{\lambda^2-4}.
\end{equation}

\noindent
This is the transcendental equation for the QBS energies of a three-equal $\delta$-function potential array. In the specific case when $\lambda=1$, Eq. 
(\ref{qbs3}) results in $kx_2 = \pi/2$ and $kx_2 = \cos^{-1}(-4/5) = 2.498$, which are shown in Figure 2 a).

\medskip

\subsection{Four-$\delta$-function potential array}

For a $N=4$ system, Eq. (\ref{d1}) reduces to

\begin{equation}\label{m4}
        {\bf \mathcal{M}}(4)= I-\frac{\imath}{2}\sum_{n_1=1}^4\lambda_{n_1}{\bf L}_{n_1} + 
         -\frac{1}{4}\sum_{\substack{n_1,n_2=1\\n_1>n_2}}^4 \lambda_{n_1}{\bf L}_{n_1}\lambda_{n_2}{\bf L}_{n_2}
         \nonumber
\end{equation} 
\begin{equation}
        +\frac{\imath}{8}\sum_{\substack{n_1,n_2,n_3=1\\n_1>n_2>n_3}}^4 
        \lambda_{n_1}{\bf L}_{n_1}\lambda_{n_2}{\bf L}_{n_2}\lambda_{n_3}{\bf L}_{n_3}+
        \frac{1}{16}\lambda_{n_1}{\bf L}_{n_1}\lambda_{n_2}{\bf L}_{n_2}\lambda_{n_3}{\bf L}_{n_3}\lambda_{n_4}{\bf L}_{n_4}.
\end{equation}

Note that $ {\bf \mathcal{M}}(4)$ is a product of  ${\bf \mathcal{M}}(2)$ for ${\bf L}_1$ and ${\bf L}_2$, and ${\bf \mathcal{M}}(2)$ for ${\bf L}_3$ and
${\bf L}_4$. Depending on specific relations between the four phases $E_i$ (equivalently ${\bf L}_i$), Eq. (\ref{m4}) will represent arrays with $N=1$, $2$
and $3$. For a genuine $N=4$ array we need the quadrilinear term ${\bf Q}={\bf L}_1{\bf L}_2{\bf L}_3{\bf L}_1\neq 0$. Using the properties of the 
$L$-matrices one can read off the following special cases:

\begin{enumerate}[i)]
  \item The effect of resonance reduces the array to an effective $N = 1$ system, if all $E_i$ (or ${\bf L}_i$) $i=1$, $2$, $3$ and $4$ are equal.

  \item  The array is reduced to an effective $N=2$ system. These can arise in $3$-ways, namely a) ${\bf L}_1={\bf L}_2={\bf L}_3$, ${\bf L}_4$, b)
         ${\bf L}_1$, ${\bf L}_2={\bf L}_3={\bf L}_4$ and c) ${\bf L}_1={\bf L}_2$, ${\bf L}_3={\bf L}_4$. 

  \item The array is reduced to a three-$\delta$-function system. For an effective $N=3$ case, one of the trilinear products;
        ${\bf \tau}_1\equiv {\bf L}_4{\bf L}_3{\bf L}_1$, ${\bf \tau}_2\equiv {\bf L}_4{\bf L}_3{\bf L}_2$, ${\bf \tau}_3\equiv {\bf L}_3{\bf L}_2{\bf L}_1$,
        ${\bf \tau}_4\equiv {\bf L}_4{\bf L}_2{\bf L}_1$ should be non-zero with the quadrilinear term $Q = 0$. Then, two adjacent $L$-matrices should be
        equal and consequently the effective strength parameter would be the sum of the original parameters ($\lambda_i$'s). In this situation we distinguish
        the following different cases:

        \begin{enumerate}
        
        \item ${\bf L}_1={\bf L}_2$, so $T_1 = T_2 =  {\bf L}_4{\bf L}_3{\bf L}_1$, with $T_3 = T_4 = 0$. The effective strength parameters at the three 
              $\delta$-function potentials are ($\lambda_1 + \lambda_2$), $\lambda_3$ and $\lambda_4$.

        \item ${\bf L}_1={\bf L}_2={\bf L}_4$, so $T_1=T_2={\bf L}_1{\bf L}_3{\bf L}_1$. The effective strength parameters are as in case (a).

        \item ${\bf L}_2={\bf L}_3$, so $T_1=T_4={\bf L}_4{\bf L}_2{\bf L}_1$, with $T_2=T_3=0$. The effective strength parameters are $\lambda_1$,
              ($\lambda_2 + \lambda_3$) and $\lambda_4$.

        \item ${\bf L}_4={\bf L}_1$, gives $T_2=T_4={\bf L}_1{\bf L}_3{\bf L}_1$ gives a particular case of case (c) and it is the same as case (b)
              with the replacement ${\bf L}_3 \rightarrow {\bf L}_2$ and $\lambda_3 \rightarrow \lambda_2$ .

        \item ${\bf L}_4={\bf L}_3$, gives $T_3=T_4={\bf L}_3{\bf L}_2{\bf L}_1$ with $T_1=T_2=0$. The effective strength parameters are $\lambda_1$, 
              $\lambda_2$ and ($\lambda_3+\lambda_4$).

        \item The choice ${\bf L}_3={\bf L}_1$ gives $T_3=T_4={\bf L}_1{\bf L}_2{\bf L}_1$, a particular case of case (e). However, this case is
              mathematically equivalent to case (d) since the $\lambda_i$'s, the potential strength parameters are not fixed. 

        \end{enumerate}

  \item A genuine four-$\delta$ potential with ${\bf Q}={\bf L}_1{\bf L}_2{\bf L}_3{\bf L}_1 \neq 0$. From Eq. (\ref{m4}) one obtains

  \begin{equation}
  \label{m42}
  ({\bf \mathcal{M}}(4))_{22} = \frac{1}{16}(z_1z_2z_3z_4 + \lambda_1\lambda_2z_3z_4E_2E_1^* + \lambda_1\lambda_3z_2^*z_4E_3E_1^* + 
  \lambda_1\lambda_4z_2^*z_3^*E_4E_1^* 
         \nonumber
  \end{equation} 
  \begin{equation}
        +\lambda_2\lambda_3z_1z_4E_3E_2^* + \lambda_2\lambda_4z_1z_3^*E_4E_2^* + \lambda_3\lambda_4z_1z_2E_4E_3^* + 
      \lambda_1\lambda_2\lambda_3\lambda_4E_4E_2E_3^*E_1^*).
  \end{equation}
  
\end{enumerate}

To find the conditions for perfect tunnelling in a genuine $N=4$ case, we choose the origin at $x_1=0$ ($E_1=1$), and we suppose that all $\delta$-potential
have the same strength ($\lambda_1 = \lambda_2 = \lambda_3 = \lambda_4 = \lambda$). We expect the existence of intrinsic QBS energies between every two
contiguous potential, which have to be all equal to allow perfect tunnelling. As these energies depend on the potential separation, we infer that all
separations are equal, meaning that $E_4 = E_2^3$ and $E_3 = E_2^2$. With this considerations Eq. (\ref{m42}) reduces to   

\begin{equation}\label{m43}
  ({\bf \mathcal{M}}(4))_{22} = \frac{1}{16}(z^4 + \lambda^2E_2(3z^2 + 2|z|^2E_2 + z^{*2}E_2^2 + \lambda^2E_2)).
\end{equation} 

\noindent
Imposing the perfect tunnelling condition $T = 1/|({\bf \mathcal{M}}(3))_{22}|^2=1$, we obtain

\begin{equation}\label{m422}
4\cos(2kx_2)+6\lambda^2\sin(kx_2) = \lambda\tan(kx_2)(2-12\cos^2(kx_2)-\lambda^2\sin^2(kx_2)),
\end{equation}

\noindent
which is the transcendental relation to find the QBS energies of a four-equal strength $\delta$-function potential array. 

It should be noted that a special situation is present in the $N = 4$ potential array; when the first two and the last two potentials have perfect tunnelling
separately. In this case, the four-$\delta$-function array can be seen as a system of two pairs of sub-arrays or cells with perfect tunnelling conditions. As
we will discuss in the next section, we expect perfect tunnelling if the QBS energies of these two potential cells are the same.

\section{Results for arbitrary N}

As we seen in the cases of specific arrays, it is always possible to use the resonance of the wavefunction with the geometry of the potential array to
effectively reduce the number of potentials, and the complexity of the problem. This reduction is allowed only when the resonance is present in a row of 
adjacent potential.

Similar to that seen in the previous section, to obtain perfect tunnelling in a $N$ equal-strength $\delta$-function potential array, we can locate them in a
series with the same distance one after the other. This assures us the matching of all the intrinsic QBS energies which will be the QBS energies of the whole
array. Another way to obtain perfect tunnelling in an $N$ $\delta$-function potential array is to separate it in cells of potentials, with perfect tunnelling
conditions by themselves, as we in Figure 3 a). We expect to have perfect tunnelling when the incident particle energy is equal to one QBS energy in
every cell. The advantage of this method is that in principle, the two cells in Figure 3 a) can be located at any distance one from each other with some
resonance in the non perfect tunnelling energies. 

\begin{figure}[h]
\begin{center}
\resizebox{14cm}{!}{\includegraphics{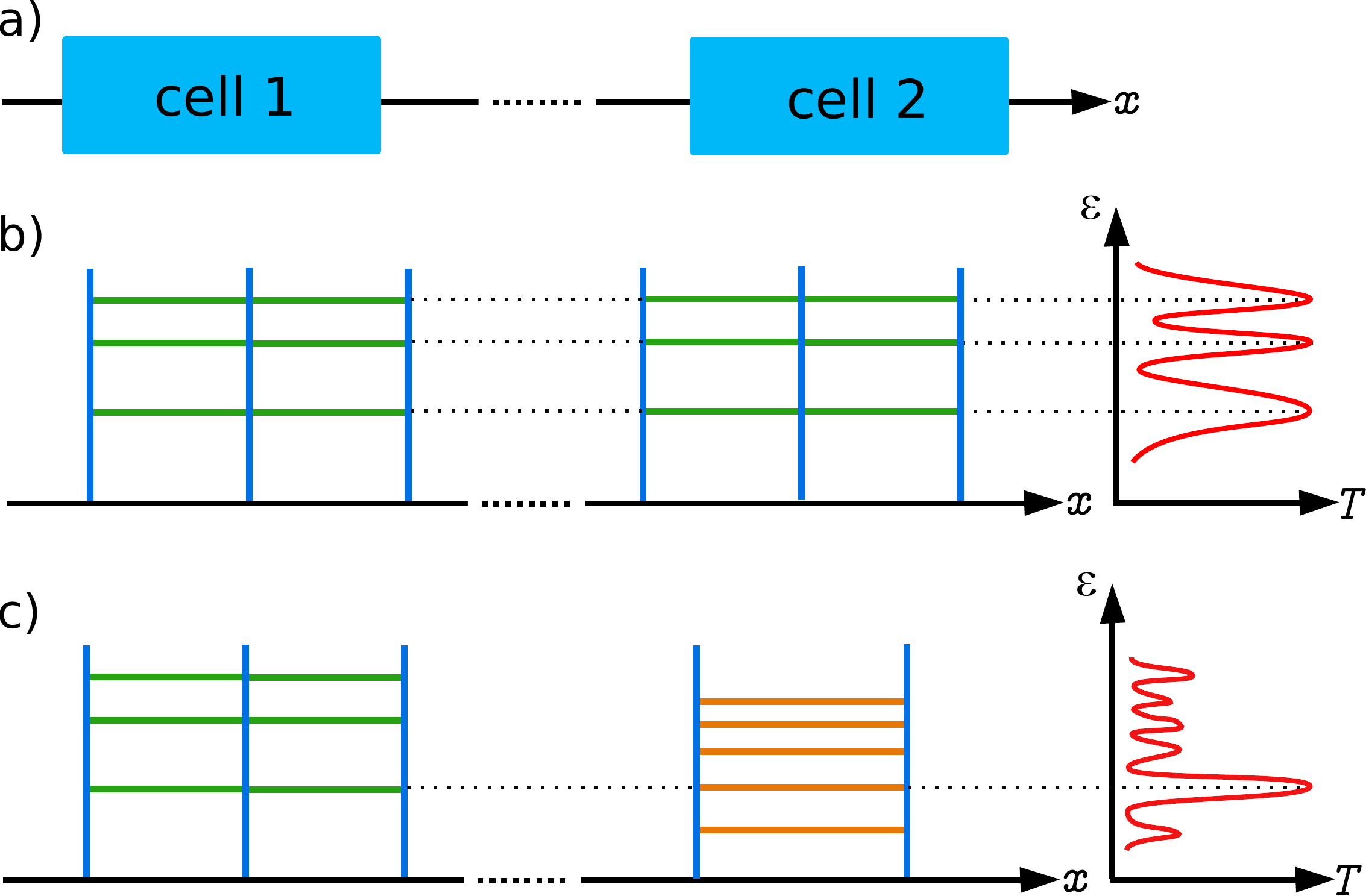}}
\caption{(Color online) Schemes of perfect tunnelling in a cells configuration. a) Two perfect tunnelling cells can be located one after the other, at any
          separation. b) Two equal cells of three $\delta$-function potentials are located one after the other. It is expected perfect tunnelling in the QBS
          energies of the cells. c) Two different cells (three and two $\delta$-function potentials) are located in series. As they have different QBS
          energies, we expect to have perfect tunnelling only when two of these energies match.}
\end{center}
\end{figure}

For example, if we locate two or more identical cells, we can expect perfect tunnelling at the incident energies which match all the QBS energies of a
single cell (see Figure 3 b)). By contrast, if the cells are different, such as three and two $\delta$-function potentials as is shown in Figure 3 c),
we only have the perfect tunnelling conditions when incident energy is equal to one QBS energy in every cell. Otherwise we can expect maximal values in the
transmitivitty but not perfect tunnelling. 

\begin{figure}[h]
\begin{center}
\resizebox{16cm}{!}{\includegraphics{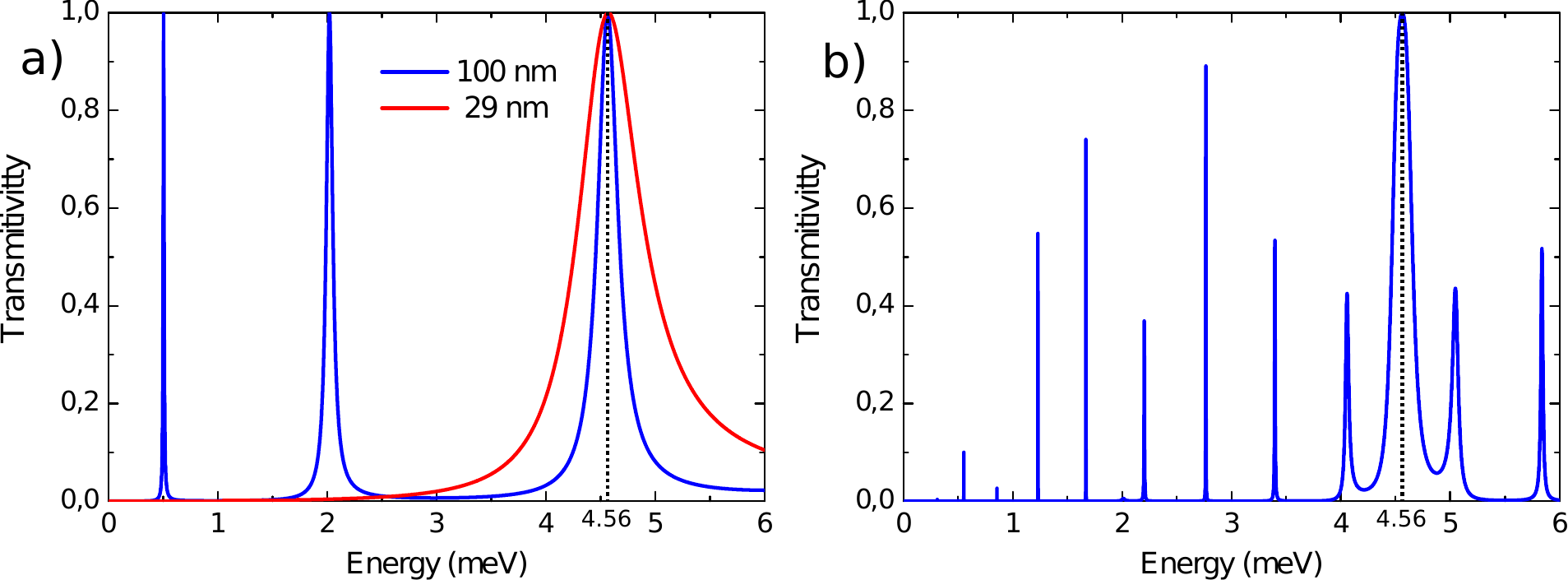}}
\caption{(Color online) Perfect tunnelling selectivity in GaAs quantum wire. a) Particle transmitivitty of two different two-$\delta$-function potential
          arrays as a function of incident energy. In blue line the tranmitivitty of two equal $\delta$-function potential barriers with $J = 2$ eV\AA ,
          separated by $100$ nm. In red line the tranmitivitty of two equal $\delta$-function potential barriers with $J = 2$ eV\AA , separated by $29$ nm.
          Matching perfect tunnelling is found at $4.56$ meV. b) Particle transmitivitty of a four-$\delta$-function potential array as a function of
          incident energy. The four potential is arranged in two cells of two-$\delta$-function potentials depicted in a) separated by $150$ nm. Perfect
          tunnelling is found at $4.56$ meV.}
\end{center}
\end{figure}

The difference in QBS energies of each cell can be used as an energy selector or filter; showing perfect tunnelling only at one defined value of incident
energy, and preventing the transmission in other energy values. To illustrate our proposal we consider a GaAs quantum wire (where the effective mass of
electron is $0.067$ times the electron mass) and locate two $\delta$-function potential barriers with equal strength ($J = 2$ eV\AA ). The transmitivitty as a
function of incident energy in this case is shown in Figure 4 a), in blue line when the potentials are located $100$ nm one to each other. The red line shows
the transmitivitty when the potentials are separated by $29$ nm. As can be seen, both cases share a QBS energy at $4.56$ meV. Now, we take these two cases and
put them together in series separated by $150$ nm, to form a four-$\delta$-function potential array. In the Figure 4 b) we show the transmitivity as a
function of the incident energy in this four-$\delta$-function potential array. Note that in this configuration the perfect tunnelling energy at $4.56$ meV
is preserved, while the other maximum values disappear. In this example the two cells behave like potential barriers for incident energies different than QBS energies. Considering the cells as potential barriers, the arbitrary separation between the two cells (in the example $150$ nm) can create resonances for different incident energies, but never perfect tunnelling, unless the two cells are equal, as it was shown in \cite{cor}. 

\section{Conclusions}

Conditions for resonant transmission are considered, in detail, for arrays with $N = 1$, $2$, $3$ and $4$. These results were applied to potential arrays with
arbitrary $N$, using a simple representation for the transmission matrix. The resonant behaviour is also the cause of the perfect tunnelling present in these
systems. In this context, we calculate the transcendental relations for QBS energies in specific arrays, using the concept of intrinsic QBS energies. For
arrays with arbitrary $N$, we propose the separation of the array into subsets or cells, whose QBS energies are related with the QBS energies of the whole system. Using the relation between intrinsic QBS energies we showed how a energy filter works.

\section*{Acknowledgments} \label{ack}

G.C.M. thanks to Consejo Nacional de Ciencia y Tecnolog\'ia of Mexico (Conacyt-Mexico) for a postdoctoral grant. This research was partially supported by
Conacyt-Mexico under grant No. 83604.


\begin{thebibliography}{40} 


\bibitem{gri}  D. J. Griffiths and C. A. Steinke, Am. J. Phys. {\bf 69}, 137 (2001). 

\bibitem{pw} W. C. Elmore and M. A. Heald, Physics of Waves, 1 st ed. (Dover, New york, 1985).

\bibitem{cra} F. S. Crawford, Am. J. Phys. {\bf 42}, 278 (1974͒).

\bibitem{san} J. V. S\'anchez-P\'erez \emph{et al.}, Phys. Rev. Lett. {\bf 80}, 5325 (1998͒).

\bibitem{pochi} P. Yeh, A. Yariv and C Hong, J. Opt. Soc. Am. {\bf 67}, 423 (1977).

\bibitem{ben} J. M. Bendickson, J. P. Dowling and M. Scalora, Phys. Rev. E {\bf 53}, 4107 (1996). 

\bibitem{va} V. B. Kazanskiy and V. V. Podlozny, Microwave Opt. Technol. Lett. {\bf 21}, 299 ͑(1999͒).

\bibitem{gri2} D. J. Griffiths and N. F. Taussig, Am. J. Phys. {\bf 60}, 883 (1992).

\bibitem{wo} H. Wu, D. W. L. Sprung and J. Martorell, J. Phys. D: Appl. Phys. {\bf 26}, 798 (1993).

\bibitem{spr} D. W. L. Sprung, H. Wu and J. Martorell, Am. J. Phys. {\bf 61}, 1118 (1993). 

\bibitem{cv} M. \v{C}vetit and L. Pi\v{c}man, J. Phys. A: Math. Gen. {\bf 14}, 379 (1981). 

\bibitem{bra} B. Gutierrez-Medina, Am. J. Phys. {\bf 81}, 104 (2013).

\bibitem{esa1} L. Esaki, Phys. Rev. {\bf 109}, 603–604 (1958).

\bibitem{esa2} R. Tsu and L. Esaki, Appl. Phys. Lett. {\bf 22}, 562 (1973).

\bibitem{esa3} L. Esaki, Y. Arakawa and M. Kitamura, Nature {\bf 464}, 31 (2010).

\bibitem{esa4} L. Esaki, Nobel Lecture (1973).

\bibitem{ngu} H. S. Nguyen \emph{et al.}, Phys. Rev. Lett. {\bf 110},  236601 (2013).

\bibitem{ro1} S. Nonoyama \emph{et al.}, Phys. Rev. B {\bf 47}, 2423 (1993).
  
\bibitem{ro2} J. Besprosvany, Phys. Rev. B {\bf 63}, 233108 (2001).
  
\bibitem{ro3} A. Agarwal and D. Sen, Phys. Rev. B {\bf 73}, 045332 (2006).
  
\bibitem{ro4} Y. G. Peisakhovich and A. A. Shtygashev, Phys. Rev. B {\bf 77}, 075327 (2008).

\bibitem{kit} C. Kittel, Introduction to Solid State Physics, 8 th ed. (Wiley, New york, 2004).

\bibitem{and} P.W. Anderson, Phys. Rev. {\bf109}, 1492 (1958).
  
\bibitem{hue} P. Ojeda, R. Huerta-Quintanilla and M. Rodriguez-Achach, Phys. Rev. B {\bf 65}, 233102 (2002).
  
\bibitem{koh} M. Kohmoto, Phys. Rev. B {\bf 34}, 5043 (1986).
  
\bibitem{si} J. Sak and B. Kramer, Phys. Rev. B {\bf 24}, 1761 (1981).

\bibitem{bos}  A.T. Costa, Jr., S. Bose and Y. Omar, Phys. Rev. Lett.~\textbf{96}, 230501 (2006).
  
\bibitem{cic1} F. Ciccarello \emph{et al.}, New J. Phys. {\bf 8}, 214 (2006).
  
\bibitem{cic2} F. Ciccarello, M. Paternostro, M. S. Kim and G. M. Palma, Phys. Rev. Lett. {\bf 100}, 150501 (2008). 

\bibitem{cor2} G. Cordourier-Maruri \emph{et al.}, Phys. Rev. A. {\bf 82}, 052313 (2010).

\bibitem{cicn2} F. Ciccarello \emph{et al.}, Phys. Rev. A {\bf 82},  030302(R) (2010).

\bibitem{cicn} F. Ciccarello \emph{et al.}, Phys. Rev. A {\bf 85}, 050305(R) (2012).

\bibitem{cit} F. Ciccarello, S. Bose and M. Zarcone, Phys. Rev. A {\bf 81}, 042318 (2010).

\bibitem{qm} A. De Pasquale, F. Ciccarello, K. Yuasa and V Giovannetti, New J. Phys. {\bf 15}, 043012 (2013).

\bibitem{white} C. T. White and T. N. Todorov, Nature {\bf 393}, 240-242 (1998).

\bibitem{balents} L. Balents and R. Egger, Phys. Rev. Lett. {\bf 85}, 3464 (2000).

\bibitem{hermelin} S. Hermelin \emph{et al.}, Nature {\bf 477}, 435 (2011).

\bibitem{mcnell} R. P. G. McNell {\it et al.}, Nature {\bf 477}, 439 (2011).
 
\bibitem{cor} G. Cordourier-Maruri, R. de Coss and V. Gupta, Mod. Phys. Lett. b {\bf 25}, 1349 (2011).

  
\end{thebibliography}
\end{document}